\documentclass[a4paper,11pt]{article}

\usepackage{jheppub} 

\usepackage[T1]{fontenc} 
\usepackage{bm,latexsym,amsmath,amssymb,amsfonts,mathtools,mathrsfs}
\title{\boldmath Vacuum decays around spinning black holes}

\author[a]{Naritaka Oshita}
\author[b]{Kazushige Ueda}
\author[c]{Masahide Yamaguchi}

\affiliation[a]{Perimeter Institute, 31 Caroline St, Waterloo, Ontario N2L 2Y5, Canada}
\affiliation[b]{Graduate School of Science Hiroshima University,1-3-1 Kagami Yama, Higashi-Hiroshima, 739-8526, Japan}
\affiliation[c]{Department of Physics, Tokyo Institute of Technology, 2-12-1 Ookayama, Meguro-ku, Tokyo 152-8551, Japan}

\emailAdd{noshita@pitp.ca}
\emailAdd{m180959@hiroshima-u.ac.jp}
\emailAdd{gucci@phys.titech.ac.jp}

\abstract{We investigate a vacuum decay around a spinning seed black hole by using the Israel junction condition and conclude that the spin of black hole would suppress a vacuum decay rate compared to that for a non-spinning case, provided that the surface of vacuum bubble has its ellipsoidal shape characterized by the Kerr geometry. We also find out that in the existence of a near-extremal black hole, a false vacuum state can be more stabilized than the case of the Coleman-de Luccia solution. A few necessary assumptions to carry the calculations are discussed.}

\begin{document} 
\maketitle
\flushbottom

\section{Introduction}
The tunneling process is one of the most novel features to characterize
quantum mechanics. A metastable state, which is a local minimum of a
potential but not a global one, is still stable at classical level, but
it can tunnel to the global minimum quantum mechanically. Such a quantum
tunneling process in a field theory was first discussed without
\cite{Kobzarev:1974cp,Coleman:1977py,Callan:1977pt} and with
\cite{Coleman:1980aw} gravity effects taken into account long times ago.
In these works, $O(4)$ symmetric bounce solutions are considered because
it was proven \cite{Coleman:1977th} that they give the least Euclidean
action without gravity, that is, in Euclidean space, though it is still
an open question whether this is robust in cases with gravity taken into
account.

As a result of a tunneling process, a bubble filled with a true vacuum
is nucleated among a false vacuum sea, which has a lot of interesting
implications to cosmology.  The nucleation of a bubble triggers the
creation of a wormhole \cite{Sato:1981bf} as well as an inflation with
open Universe \cite{Gott:1982zf}, which has attracted a renewed interest
in the context of string landscape
\cite{Susskind:2003kw,Freivogel:2004rd}.  The collisions of nucleated
bubbles are important source of the generation of gravitational waves
(see e.g. \cite{Cai:2017cbj,Mazumdar:2018dfl} for recent review).

The quantum tunneling process has also been paid particular attention
in the context of Higgs physics. Higgs particle was finally found in LHC
experiments several years ago \cite{Aad:2012tfa,Chatrchyan:2012xdj} and
its self-coupling suggests that our vacuum might be metastable and can
decay into the true vacuum after taking its running into account
\cite{Chatrchyan:2012xdj,Sher:1988mj,Arnold:1989cb,Altarelli:1994rb,Espinosa:1995se,Casas:1996aq,Hambye:1996wb,Isidori:2001bm,Espinosa:2007qp,Ellis:2009tp,Bezrukov:2012sa,Bednyakov:2015sca,EliasMiro:2011aa,Degrassi:2012ry,Buttazzo:2013uya,Branchina:2013jra,Greenwood:2008qp}. But, very
fortunately, the detailed calculations show that the life time of our
vacuum is much longer than the cosmic age (see
e.g. \cite{Chigusa:2017dux,Chigusa:2018uuj} for the recent estimate).
However, these estimates are based on $O(4)$ symmetric bounce solution
assuming not only spatial isotropy but also its homogeneity, which can
be broken if there is an impurity. Hiscock first discussed the effects
of a non-spinning black hole (BH) on a vacuum decay and show that it can
enhance the decay rate \cite{Hiscock:1987hn}. Later many authors refined
his analysis, and, the importance of the change of the Bekenstein-Hawking
entropy of the BH was pointed out by Gregory \textit{et al.}
\cite{Gregory:2013hja}. Two of the present authors with
Yamada also suggested that not only BHs but also compact objects such as
monopoles, Q-balls, oscillons, boson stars (including axion stars),
gravastars, neutron stars can catalyze a vacuum decay and significantly
enhance its decay rate because of the absence of BH entropies \cite{Oshita:2018ptr}. However,
BHs and compact objects are still assumed to be spherically symmetric
(spatially isotropic) and then only spherically symmetric bounce
solutions have been considered so far.

In this paper, as far as we know, for the first time, we discuss the
effects of a {\it non-spherical} compact object on vacuum decay through
quantum tunneling. As such a concrete example, we investigate the
nucleation of a vacuum bubble around a spinning BH by
assuming that the vacuum decay does not change the angular momentum and mass of
the seed BH. We consider the first order phase transition from a false
vacuum state with zero vacuum energy to a true vacuum state.  We assume
that the typical shape of the nucleated vacuum bubble would be
determined by the angular components of
metric, which is a natural extension of the case of vacuum decays around
Schwarzschild BHs \cite{Gregory:2013hja}. That is, the nucleated vacuum
bubble is assumed to be ellipsoidal throughout this paper.

The organization of this paper is as follows. In the next section, we
briefly review the effects of spherically symmetric (non-rotating) BHs
on vacuum decay based on \cite{Hiscock:1987hn,Gregory:2013hja}. We derive the extrinsic curvature of a dynamical ellipsoidal boundary in Kerr spacetime and Kerr-anti-de Sitter (Kerr-AdS) spacetime in Sec. \ref{subsec:excurve} and the second Israel junction
condition on the boundary of the wall around the black hole in Sec. \ref{subsec:second_Israel}
in order to obtain the Euclidean solution of the spheroidal thin-wall bubbles. In Sec. \ref{subsec:class} we then classify the obtained Euclidean solutions to two classes: physically meaningful solutions and unphysical/unrealistic solutions. In Sec. \ref{subsec:euclidean_action}, we give the Euclidean action, which is the exponent of the decay rate around a Kerr BH, and discuss how much its decay rate is changed
compared to the case without impurities \cite{Coleman:1980aw} and the case with a spherically
symmetric (non-rotating) BH \cite{Hiscock:1987hn,Gregory:2013hja}. The final section is devoted to conclusions. We use the natural units $c=\hbar=1$ throughout the paper and $G = \ell_{\text{Pl}}^2 = M_{\text{Pl}}^{-2}$, where $M_{\text{Pl}}$ and $\ell_{\text{Pl}}$ are the Planck mass and Planck length, respectively.

\section{Brief review of vacuum decay around a non-rotating BH}
\label{sec:review}

Here we briefly review the vacuum decay around a non-rotating BH, pioneered by Hiscock \cite{Hiscock:1987hn}. He calculated the Euclidean action of a vacuum bubble surrounding a static BH at the origin by imposing the thin-wall approximation, and obtained two primary results that the Euclidean action of a vacuum bubble in the existence of the seed BH is always less than the corresponding Coleman-de Luccia (CDL) bubble action \cite{Coleman:1980aw}, and that there is the maximum mass of the seed BH, below which the classical Euclidean solution exists. Gregory {\it et al.} \cite{Gregory:2013hja}, however, improved the calculation of the Euclidean action in the existence of the BH by properly taking the conical singularities into account, and it was shown that the resulting action can be larger than the CDL action only when the background spacetime is close to the Nariai limit \cite{nariai1,nariai2}. Other than this point, both conclusions in \cite{Hiscock:1987hn} and \cite{Gregory:2013hja} are qualitatively consistent. In the following, a brief review of the Euclidean solution around a BH, based on \cite{Hiscock:1987hn,Gregory:2013hja}, is presented.

\subsection{Thin-wall vacuum bubble around a BH}
In the semi-classical approximation, a vacuum decay rate, $\Gamma$, can be estimated by the exponential of an on-shell Euclidean action $S_{\text{E}}$, and so one has to investigate the (classical) Euclidean dynamics of a vacuum bubble in order to calculate $S_{\text{E}}$. We restrict ourselves to the case of thin-wall vacuum bubble throughout the manuscript, which allows us to use the Israel junction condition to analytically investigate the dynamics of the thin-wall bubble.

Let us derive the junction condition between the interior and exterior spacetimes whose metrics, $g^{(+)}_{\mu \nu}$ and $g^{(-)}_{\mu \nu}$, are given by
\begin{align}
ds^2 &= g_{\mu \nu}^{(\pm)} dx^{\mu}_{\pm} dx^{\nu}_{\pm} = - f_{\pm} dt_{\pm}^2 + \frac{dr_{\pm}^2}{f_{\pm}} + r_{\pm}^2 d\Omega_{\pm}^2,\label{metric_review}\\
f_{\pm} (r) &= 1-\frac{2GM_{\pm}}{r} + H_{\pm}^2 r^2,
\end{align}
respectively, where $d\Omega_{\pm}^2 \equiv d\theta_{\pm}^2 + \sin^2{\theta_{\pm}} d\phi_{\pm}^2$. Respecting the symmetry of (\ref{metric_review}), we will investigate a Euclidean bubble with $O(3)$-symmetry, and its surface, $\Sigma_{{\cal W} \pm}$, is given by
\begin{equation}
\Sigma_{{\cal W} \pm} = \left\{ (t_{\pm}, r_{\pm}, \theta_{\pm}, \phi_{\pm}) | F_{\pm} (t_{\pm}, r_{\pm}) = r_{\pm}-R (\tau (t_{\pm}))=0 \right\},
\label{surface_shell_review}
\end{equation}
where $\tau$ is the proper time on the wall. The induced metrics on $\Sigma_{{\cal W} \pm}$ , $h^{(\pm)}_{ab}$, is given by
\begin{equation}
ds^2 = h^{(\pm)}_{ab} dx^a dx^b= \left( -f_{\pm} (R (\tau)) \dot{t}_{\pm}^2 + \frac{\dot{r}_{\pm}^2}{f_{\pm} (R(\tau))} \right) d\tau^2 + R^2(\tau) d\Omega_{\pm}^2 =  -d\tau^2 + R^2(\tau) d\Omega_{\pm}^2,
\label{induced_review}
\end{equation}
where we used $X_{\pm}^{\mu} X_{\pm \mu} = -1$ and $X_{\pm}^{\mu} \equiv (\dot{t}_{\pm}, \dot{R},0,0)$ is the four velocity of the wall.
The first Israel junction condition requires the continuity between the interior and exterior induced metrics on the wall, and from (\ref{induced_review}), one can read that $h^{(+)}_{ab} = h^{(-)}_{ab}$ obviously holds.

\begin{figure}[h]
\centering
    \includegraphics[width=0.6 \textwidth]{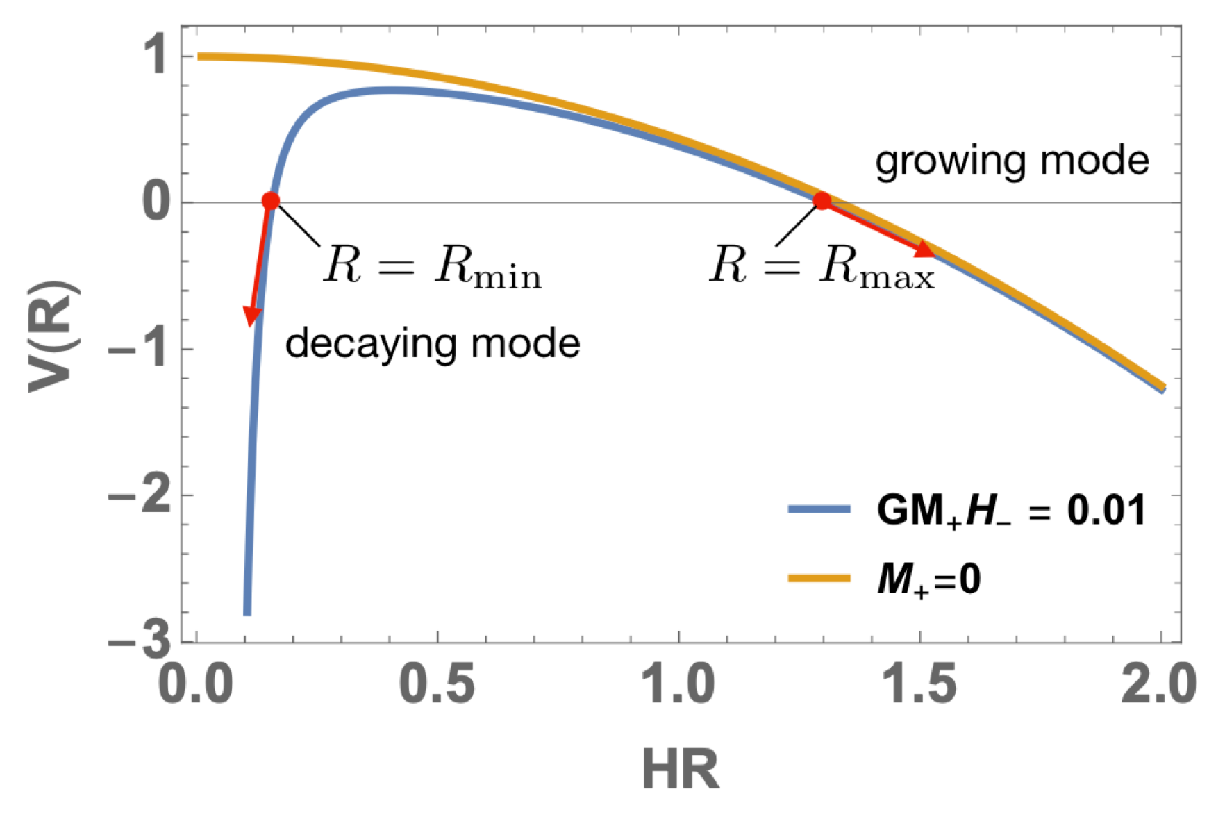}
\caption{Plot of the effective potential $V(R)$ with $H_+ =0, M_-=0, H_-=H > 0, \Sigma = H/2$, and $M_+ = 0.01/(G H)$.
}
\label{potential_review}
\end{figure}

The second Israel junction condition is given by
\begin{equation}
K_{+ ab} - K_{- ab} = - 8 \pi G \left(S_{ab} -\frac{1}{2} h_{ab} S \right), 
\label{israel_review}
\end{equation}
where $K_{\pm ab}$ is the extrinsic curvature on $\Sigma_{{\cal W} \pm}$, $S_{ab}$ is the energy momentum tensor of the wall, and $S \equiv \text{Tr}(S_{ab})$. Throughout this section, we assume $S_{ab} = -\sigma h_{ab}$. The extrinsic curvature is given by $K_{\pm ab} = \nabla_a n_{\pm b}$, where $n_{\pm \mu}$ is the unit vector normal to $\Sigma_{{\cal W} \pm}$:
\begin{equation}
n_{\pm \mu} = \frac{\partial_{\mu} F_{\pm}}{\sqrt{|g^{(\pm)} {}^{\mu \nu} \partial_{\mu} F_{\pm} \partial_{\nu} F_{\pm}|}}.
\label{unit_normal_review}
\end{equation}
Using (\ref{surface_shell_review}), one obtains
\begin{equation}
n_{\pm \mu} = (- \dot{R}, \dot{t}_{\pm},0,0),
\end{equation}
and the ($\theta, \theta$)-component of the extrinsic curvature is
\begin{equation}
K_{\pm \theta \theta} = R f_{\pm} \dot{t}_{\pm}.
\end{equation}
One can obtain the energy conservation law of the wall from the ($\theta, \theta$)-component of the second junction condition, which has the form of
\begin{equation}
\frac{f_+ \dot{t}_+}{R} - \frac{f_- \dot{t}_-}{R} = -4 \pi G \sigma \equiv -\Sigma.
\label{junction_th_th_review}
\end{equation}
Using $X_{\pm}^{\mu} X_{\pm \mu} = -1$, the energy-conservation law is obtained as
\begin{align}
&\dot{R}^2 + V(R) = 0, \label{energy_conservation}\\
&V(R) \equiv f_- - \frac{1}{4 \Sigma^2 R^2} \left( \frac{2G (M_+ - M_-)}{R} + (H_-^2 -H_+^2 +\Sigma^2) R^2 \right)^2,
\end{align}
where $V(R)$ is an effective potential which governs the position of the wall. As an example, let us consider the vacuum decay from zero-energy vacuum state ($H_+ = 0$) to an AdS vacuum ($H_- =H>0$) around a BH with mass $M_+$ while there is no remnant BH ($M_- = 0$). In this case, the effective potential has a concave shape which has its maximum at a finite radius, but if there is no seed BH, which is nothing but the CDL case, $V(R)$ is a monotonically decreasing function (Fig. \ref{potential_review}). The initial size of vacuum bubble, $R_0$, is given by the root of $V(R_0) = 0$, and in the former case, there exist two solutions. We have the small and large-size bubbles, whose radius are denoted as $R_{\text{min}}$ and $R_{\text{max}}$, correspond to the decaying and growing modes, respectively. On the other hand, for the CDL solution, one has the unique initial radius of $R_0 = 2\Sigma/|H^2-\Sigma^2|$, and the bubble expands after its nucleation.

\subsection{$O(3)$ Euclidean solution}
\begin{figure}[h]
\centering
    \includegraphics[width=0.75 \textwidth]{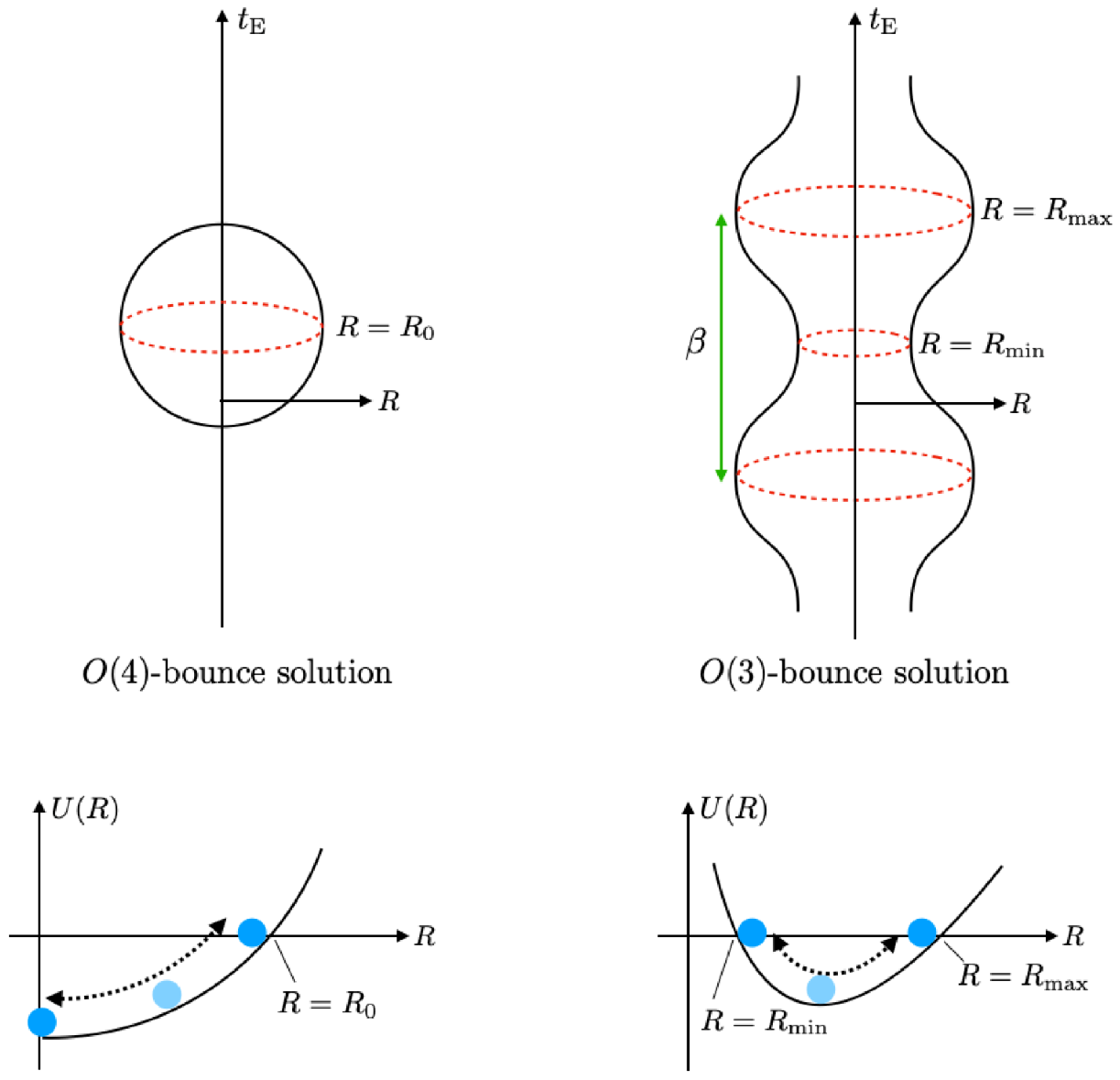}
\caption{Schematic pictures showing the $O(4)$ and $O(3)$-bounce solutions.
}
\label{euclidean_pic_1}
\end{figure}

\begin{figure}[h]
 \centering
    \includegraphics[width=0.9 \textwidth]{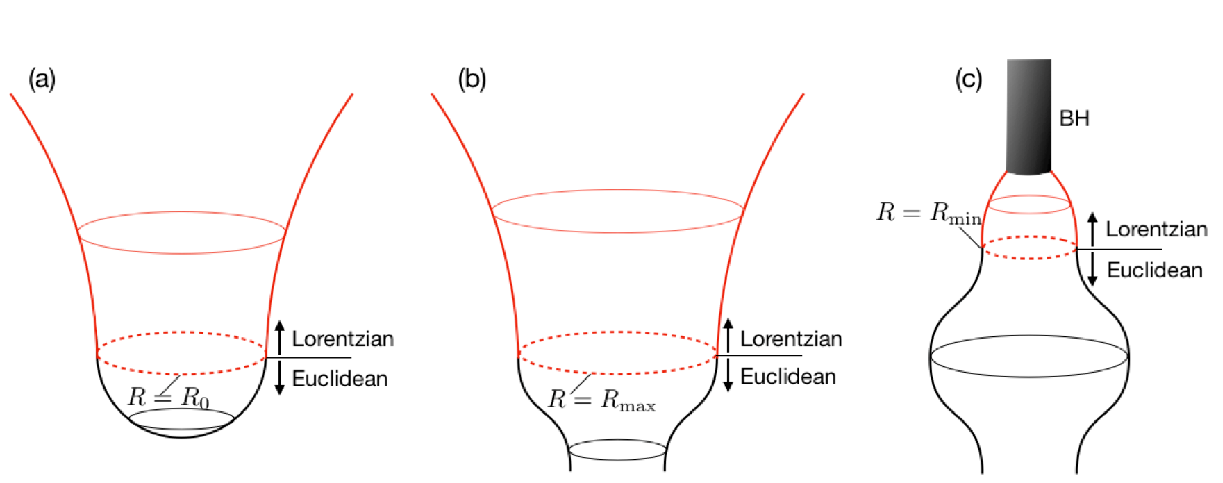}
\caption{Schematic pictures of analytic continuation between the Lorentzian and Euclidean solutions for (a) $O(4)$ bounce and (b,c) $O(3)$ bounce. The growing and decaying modes are shown in (b) and (c), respectively.
}
\label{euclidean_pic_2}
\end{figure}

\begin{figure}[h]
 \centering
    \includegraphics[width=0.7 \textwidth]{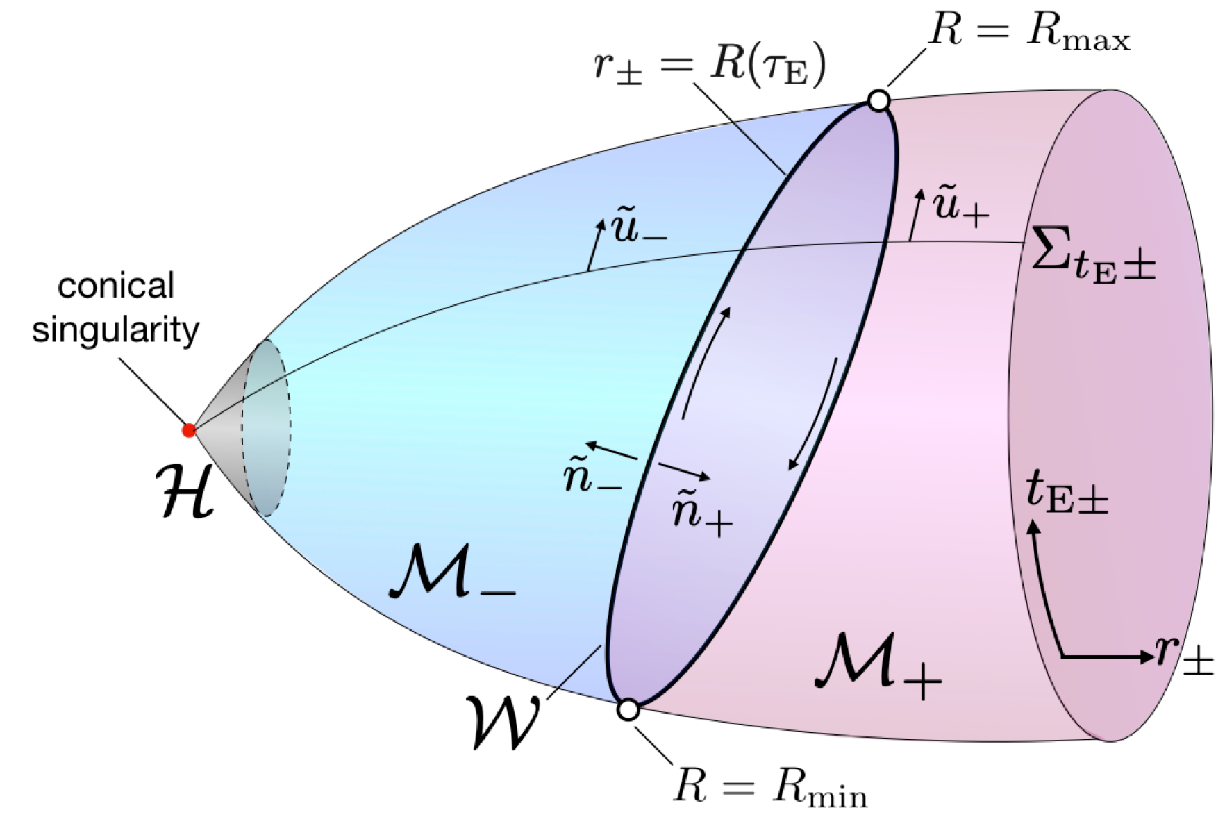}
\caption{Schematic picture showing the decomposition of the $O(3)$-Euclidean manifold in the existence of the thin-wall vacuum bubble and the BH horizon.
}
\label{picture}
\end{figure}

Let us consider the corresponding Euclidean solution with the above setup, which is obtained by implementing the Wick rotation\footnote{In this subsection, a dot denotes the derivative with respect to $\tau_{\text{E}}$.}, $t_{\pm} \to -i t_{\text{E} \pm}$ and $\tau \to -i \tau_{\text{E}}$, in (\ref{energy_conservation}). This gives an oscillatory solution with its period $\beta$ since the Euclidean wall is governed by the potential of $U(R) \equiv - V(R)$ as is shown in Fig. \ref{euclidean_pic_1}. In this case, one has two points at which the analytic continuation between the Euclidean and Lorentzian solutions is well-defined, i.e., at $R= R_{\text{min}}$ and $R=R_{\text{max}}$, although the CDL solution has only one point at which one can implement the analytic continuation. Depending on which point is chosen for the analytic continuation, a nucleated bubble wall may collapse into the seed BH or expand to fill the Universe (see Fig. \ref{euclidean_pic_2}).
The Euclidean action of $O(3)$-instanton is obtained by
\begin{equation}
S_{\text{E}} = - \int d^4x \sqrt{g_{\text{E}}} \left(\frac{1}{16 \pi G} {\cal R}^{(\text{E})} + {\cal L}_m^{(\text{E})} \right)
= - \oint d\tau_{\text{E}} \int d^3 x \sqrt{g_{\text{E}}} \left(\frac{1}{16 \pi G} {\cal R}^{(\text{E})} + {\cal L}_m^{(\text{E})} \right),
\label{euclidean_review}
\end{equation}
where $g_{\text{E}}$ is the determinant of the Euclidean metric, and ${\cal R}^{(\text{E})}$ and ${\cal L}_m^{(\text{E})}$ are the Euclidean Ricci scalar and Euclidean Lagrangian density, respectively. The line integration in (\ref{euclidean_review}) represents the integration by one period of the oscillatory Euclidean motion of the wall, denoted by $\beta$. The Euclidean manifold, accommodating a remnant black hole and oscillating Euclidean vacuum bubble, can be divided into four parts: the region near the BH horizon ${\cal H}$, the interior and exterior of the wall, ${\cal M}_-$ and ${\cal M}_+$, respectively, and the wall ${\cal W}$,
\begin{equation}
S_{\text{E}} = S_{\cal H} + S_{\cal M_-} +S_{\cal M_+} + S_{\cal W}.
\end{equation}
Carefully treating a conical singularity in the Euclidean manifold, one can obtain (see the Appendix A in \cite{Gregory:2013hja} for the details and its generalization to the case of an axisymmetric Euclidean metric is presented in the Appendix of this paper)
\begin{equation}
S_{\cal H} = - \frac{A}{4G},
\end{equation}
where $A$ is a horizon area. When the matter field is a canonical scalar field, one obtains
\begin{equation}
S_{\cal W} = - \int_{\cal W} d^3 x \sqrt{h_{\text{E}}} \int_{R-0}^{R+0} dr {\cal L}_{m} \simeq \int_{\cal W} d^3 x \sqrt{h_{\text{E}}} \int_{R-0}^{R+0} dr \sigma \delta (r-R(\tau_{\text{E}})) = \int_{\cal W} d^3 x \sqrt{h_{\text{E}}} \sigma,
\end{equation}
where $h_{\text{E}}$ is the determinant of the Euclidean induced metric $h_{\text{E} ab}$ and we used $\displaystyle {\cal L}_m^{(\text{E})} = - \frac{1}{2} \partial^{\mu} \phi \partial_{\mu} \phi - V_{\phi} (\phi) \simeq -\sigma \delta (r-R (\tau_{\text{E}}))$ in the vicinity of the wall.
The interior and exterior Euclidean actions should be accompanied by the Gibbons-Hawking-York (GHY) boundary terms since we implicitly introduced the boundaries to separate the Euclidean manifold into four parts:
\begin{equation}
S_{\cal M_{\pm}} = - \int_{\cal M_{\pm}} d^4 x \sqrt{g_{\text{E}\pm}} \left( \frac{1}{16 \pi G} {\cal R}^{(\text{E})}+ {\cal L}_m^{(\text{E})} \right) + \frac{1}{8 \pi G} \int_{\partial {\cal M}_{\pm}} d^3 x \sqrt{h_{\text{E}}} \tilde{K}_{\text{E} \pm},
\end{equation}
and implementing the ADM decomposition, one obtains
\begin{align}
\begin{split}
S_{{\cal M}_{\pm}} &= - \frac{1}{16 \pi G} \oint dt_{\text{E} \pm} \int_{\Sigma_{t_{\text{E} \pm}}} d^3 x \sqrt{g_{\text{E} \pm}} \left( {}^3 {\cal R}- \tilde{K}_{\text{E}}^2 +\tilde{K}_{\text{E} ab} \tilde{K}_{\text{E}}^{ab} + {16 \pi G} {\cal L}_m^{(\text{E})} \right)\\
&+ \frac{1}{8 \pi G} \int_{\cal W} d^3x \sqrt{h_{\text{E}}} \tilde{K}_{\text{E} \pm} + \frac{1}{8 \pi G} \int_{\cal W}d^3x \sqrt{h_{\text{E}}} \tilde{n}_{\pm \mu} \tilde{u}_{\pm}^{\nu} \nabla_{\nu} \tilde{u}_{\pm}^{\mu},
\end{split}
\label{euclid_pm_review}
\end{align}
where ${}^3 {\cal R}$ is the three-curvature, $\tilde{K}_{{\text{E}}ab}$ is the Euclidean extrinsic curvature, $\tilde{u}_{\pm}^{\mu}$ is a unit vector normal to $\Sigma_{t_{\text{E} \pm}}$ and $\tilde{n}_{\pm \mu}$ is an inward pointing unit vector normal to ${\cal W}$ (Fig. \ref{picture}), which means that $K_{\pm}$ in the Israel junction condition is related to the extrinsic curvature in the GHY term as $K_{\pm} = \pm \tilde{K}_{\text{E} \pm}$. The first term in (\ref{euclid_pm_review}) vanishes due to the Hamiltonian constraint, provided that the interior and exterior system is static. The unit vectors $\tilde{u}^{\mu}$ and $\tilde{n}_{\mu}$ have the forms of
\begin{align}
\tilde{u}^{\mu}_{\pm} &= (1/\sqrt{f_{\pm}},0,0,0),\\
\tilde{n}_{\pm \mu} &= (\mp \dot{R}, \pm \dot{t}_{\text{E} \pm},0,0),
\end{align}
and so $S_{{\cal M}_+} + S_{{\cal M}_-}$ reduces to
\begin{align}
\begin{split}
S_{{\cal M}_+} + S_{{\cal M}_-} &= \frac{1}{8 \pi G} \int_{\cal W} d^3 x \sqrt{h_{\text{E}}} (\tilde{K}_{\text{E} +} + \tilde{K}_{\text{E}-}) \\
&+ \frac{1}{8 \pi G} \int_{\cal W} d^3 x \sqrt{h_{\text{E}}} (\tilde{n}_{+ b} \tilde{u}_{+}^{t_{\text{E}}} \Gamma^{b}_{+{t_{\text{E}}} {t_{\text{E}}}} \tilde{u}_{+}^{t_{\text{E}}} + \tilde{n}_{- b} \tilde{u}_{-}^{t_{\text{E}}} \Gamma^{b}_{-{t_{\text{E}}} {t_{\text{E}}}} \tilde{u}_{-}^{t_{\text{E}}})\\
&= \frac{1}{8 \pi G} \int_{\cal W} d^3 x \sqrt{h_{\text{E}}} (-12 \pi G \sigma) - \frac{1}{16 \pi G} \int_{\cal W} d^3 x \sqrt{h_{\text{E}}} (f'_+ \dot{t}_{\text{E} +} - f'_- \dot{t}_{\text{E} -})\\
&=- \frac{3}{2} \int_{\cal W} d^3 x \sqrt{h_{\text{E}}} \sigma - \frac{1}{16 \pi G} \int_{\cal W} d^3 x \sqrt{h_{\text{E}}} (f'_+ \dot{t}_{\text{E} +} - f'_- \dot{t}_{\text{E} -}),
\end{split}
\end{align}
where we used the second Israel junction condition (\ref{israel_review}) and $\Gamma^{r}_{\pm t_{\text{E}} t_{\text{E}}} = - f_{\pm} f_{\pm}'/2$. Then, the total Euclidean action reduces to
\begin{align}
S_{\text{E}} &= - \frac{1}{2} \int_{\cal W} d^3 x \sqrt{h_{\text{E}}} \sigma - \frac{1}{16 \pi G} \int_{\cal W} d^3 x \sqrt{h_{\text{E}}} (f'_+ \dot{t}_{\text{E} +} - f'_- \dot{t}_{\text{E} -}) - \frac{A}{4G},\\
&= \frac{1}{16 \pi G} \int_{\cal W} d^3 x \sqrt{h_{\text{E}}} \left[ \left(\frac{2 f_{+}}{R}-f'_{+} \right) \dot{t}_{\text{E} +} - \left(\frac{2 f_{-}}{R}-f'_{-} \right) \dot{t}_{\text{E} -} \right] - \frac{A}{4G},
\label{euclidean_final_review}
\end{align}
Both Euclidean actions of growing and decaying modes are given by (\ref{euclidean_final_review}). Although our interest is the growing mode of vacuum bubble around the BH, we should note that the nucleation of collapsing vacuum bubble would occur with the same probability.

\section{Vacuum decay around a Kerr BH}
In this section, we extend the previous $O(3)$-Euclidean analysis around a static BH to the case of a Kerr BH that may lead to an axisymmetric Euclidean solution. As far as we know, an extension to a non-spherical vacuum bubble is attempted for the first time, and it is a very interesting question if the spin of a BH can more efficiently promote vacuum decay or not. The primary assumption to carry the calculation presented here is that a nucleated bubble has its ellipsoidal thin wall, whose exact shape is determined by the angular component of the Kerr metric. Here we consider the vacuum transition from the zero-energy vacuum state to an AdS vacuum state, and our analysis can be easily extended to other cases of vacuum decay (e.g., Kerr-dS $\to$ Kerr-AdS).

\subsection{Extrinsic curvature and the first Israel junction condition}
\label{subsec:excurve}
In this subsection, we calculate the extrinsic curvature of the ellipsoidal bubble wall.
We use the Boyer Lindquist (BL) coordinate to describe the Kerr-AdS spacetime:
\begin{align}
\begin{split}
&ds^2=\tilde{g}_{\mu \nu}(M,a,H) dx^{\mu} dx^{\nu}\\
&=-{\Delta_{H} \over\rho^2}(dt-{a\over\Sigma_{H}} \sin^2{\theta}d\phi)^2+{\rho^2\over\Delta_{H}}dr^2
+ {\rho^2\over\Delta_\theta}d\theta^2+{\Delta_\theta \sin^2{\theta}\over\rho^2}(a dt-{{r^2+a^2}\over\Sigma_{H}}d\phi)^2,
\end{split}
\label{KerrAdS}
\end{align}
where
\begin{align}
\Delta_{H}&\equiv (r^2+a^2)(1+r^2H^{2})-r_sr,\\
r_s &\equiv 2GM,\\
\Sigma_{H}&\equiv 1-a^2H^{2}, \\
\Delta_\theta&\equiv 1-a^2H^{2}\cos^2{\theta},\\
\rho^2&\equiv r^2+a^2\cos^2{\theta}.
\end{align}
Using the transformation $dt \to d\lambda$ and $d\phi \to d\psi - \tilde{g}_{\phi t}/\tilde{g}_{\phi \phi} d\lambda$, one can diagonalize this metric as
\begin{eqnarray}
ds^2 =g_{\lambda \lambda} d\lambda^2+g_{rr} dr^2+ g_{\theta \theta} d\theta^2
+g_{\psi \psi} d\psi^2,
\label{Kerrdiag_2}
\end{eqnarray}
where $g_{\lambda \lambda} \equiv \tilde{g}_{tt}-\tilde{g}_{\phi t} {}^{2}/\tilde{g}_{\phi\phi}$, $g_{rr} \equiv \tilde{g}_{rr}$, $g_{\theta \theta} \equiv \tilde{g}_{\theta \theta}$, and $g_{\psi \psi} \equiv \tilde{g}_{\phi \phi}$. This diagonalized metric is called the zero angular momentum observer (ZAMO) metric. Let us consider a vacuum decay from the zero-vacuum energy state with a spinning seed BH of $M=M_+$ to an AdS vacuum state ($H=H_- > 0$) with a remnant BH of $M = M_-$. In this case, the exterior and interior metrics, $g^{(+)}_{\mu \nu}$ and $g^{(-)}_{\mu \nu}$, are given by
\begin{align}
g^{(+)}_{\mu \nu} &= g_{\mu \nu} (M=M_+, a = a_+, H=0),\\
g^{(-)}_{\mu \nu} &= g_{\mu \nu} (M=M_-, a = a_-, H=H_-).
\end{align}
Let us introduce the vacuum bubble wall at which the interior metric is matched with the exterior one, and then determine the normal vectors on the bubble wall to calculate the interior and exterior extrinsic curvatures on the wall. To this end, we use the ZAMO metric, $ds^2 = g^{(\pm)}_{\mu \nu} dx^{\mu}_{\pm} dx^{\nu}_{\pm}$, and assume that the bubble wall is a time-like hypersurface given by
\begin{equation}
\Sigma_{{\cal W} \pm} = \left\{ (\lambda_{\pm}, r_{\pm}, \theta_{\pm}, \psi_{\pm}) | F (\lambda_{\pm}, r_{\pm}) = r_{\pm}-R (\tau (\lambda_{\pm}))=0 \right\},
\label{surface_shell}
\end{equation}
where $\tau$ is the proper time of the wall. In this assumption, the surface of the wall at $\tau = \text{const.}$ is covered by two coordinates, $\theta_{\pm}$ and $\psi_{\pm}$, and the induced metric on the wall is given by
\begin{align}
\begin{split}
d \tilde{s}^2 &= \left[ g_{\lambda \lambda}^{(\pm)} \dot{\lambda}_{\pm}^2 + g_{rr}^{(\pm)} \dot{R}^2 \right] d \tau^2 + g_{\theta \theta}^{(\pm)} d\theta^2_{\pm} + g_{\psi \psi}^{(\pm)} d\psi_{\pm}^2\\
&= - d\tau^2 + g_{\theta \theta}^{(\pm)} d\theta^2_{\pm} + g_{\psi \psi}^{(\pm)} d\psi^2_{\pm} \equiv h_{a b}^{(\pm)} dx_{\pm}^a dx_{\pm}^b,
\end{split}
\end{align}
where we used $g_{\lambda \lambda}^{(\pm)} \dot{\lambda}_{\pm}^2 + g_{rr}^{(\pm)} \dot{R}^2 = -1$, a dot denotes the derivative with respect to $\tau$, and italic indices denote $\tau$, $\theta$, or $\psi$.
The Israel junction condition is valid only when $h_{ab}^{(+)} = h_{ab}^{(-)}$ holds, which requires $g_{\theta \theta}^{(+)} = g_{\theta \theta}^{(-)}$ and $g_{\psi \psi}^{(+)} = g_{\psi \psi}^{(-)}$. Although in general this does not hold between the Kerr and Kerr-AdS metrics, one finds that this approximately holds in the case of $a_+ = a_-$, $M_+ = M_-$, and $a_{\pm}^2 \ll l^2 \equiv H^{-2}$. In the following we omit the subscripts of $+$ and $-$ in $a_{\pm}$ and $M_{\pm}$. Ignoring the term of the order of $a^2/ l^2$, the induced metric reduces to
\begin{align}
h^{(-)}_{\theta \theta} \simeq (R^2+a^2\cos^2{\theta}) =h^{(+)}_{\theta \theta},\\
h^{(-)}_{\psi \psi} \simeq \sin^2{\theta}\biggl[(R^2+a^2)+{r_s R a^2\sin^2{\theta}\over R^2 +a^2\cos^2{\theta}}\biggr] =h^{(+)}_{\psi \psi},
\label{induced_met}
\end{align}
and in the following we define $h_{ab} \equiv h_{ab}^{(\pm)}$, $g_{\theta \theta} \equiv g_{\theta \theta}^{(\pm)}$, and $g_{\psi \psi} \equiv g_{\psi \psi}^{(\pm)}$. We should emphasize that the shape of the wall is no longer spherical unlike the case of $a = 0$. From (\ref{induced_met}) we can read that $h_{\theta \theta}$ and $h_{\psi \psi}$ are spheroidal, and the length of corresponding major and minor axes of the wall, $L_a$ and $L_b$, are given by \cite{Kashargin:2011fg}
\begin{align}
L_a &\simeq \sqrt{a^2 + R^2},\\
L_b &\simeq R,
\end{align}
respectively. In the following, we consider the case of $a^2 \ll l^2$, $a_+ = a_-$, and $M_+ = M_-$, for which the first Israel junction condition, i.e., $h_{ab}^{(+)} = h_{ab}^{(-)}$, holds. Note that the effect of vacuum energy still remains in $g_{\lambda \lambda}^{(-)}$ and $g_{rr}^{(-)}$ because we keep the terms of the order of ${\cal O} (R^2 /l^2)$ in the interior metric. We next introduce unit vectors normal to $\Sigma_{{\cal W} \pm}$, $n^{(\pm)}_a$, by which one can calculate the interior and exterior extrinsic curvatures as we reviewed in Sec.{\ref{sec:review}}. 
From (\ref{unit_normal_review}) and (\ref{surface_shell}), one obtains the unit normal vectors
\begin{align}
n_{\pm \lambda}&= -\epsilon_{\pm} \dot{R}_\pm\sqrt{-g^{(\pm)}_{\lambda\lambda}g^{(\pm)}_{rr}}, \\
n_{\pm r}&= \epsilon_{\pm} \dot{\lambda}_\pm\sqrt{-g^{(\pm)}_{\lambda\lambda}g^{(\pm)}_{rr}},
\end{align}
where $\epsilon_{\pm} = +1$ or $-1$, and we should take $\epsilon_{\pm} = +1$ so that it points towards increasing $r_{\pm}$ for $\dot{\lambda}_{\pm} > 0$. We compute the extrinsic curvature inside and outside of the wall and obtain
\begin{align}
K^{(\pm)}_{\theta\theta} &= \nabla_{\theta} n_{\pm \theta} = {g_{\theta\theta,r}\over2g^{(\pm)}_{rr}}n_{\pm r}
= \dot{\lambda}_\pm {g_{\theta\theta,r}\over2g^{(\pm)}_{rr}} \sqrt{-g^{(\pm)}_{\lambda\lambda}g^{(\pm)}_{rr}} \equiv \dot{\lambda}_{\pm} A_{\pm}(R, \theta), \\
K^{(\pm)}_{\psi\psi}&= \nabla_{\psi} n_{\pm \psi} ={g_{\psi\psi,r}\over2g^{(\pm)}_{rr} }n_{\pm r}
=\dot{\lambda}_\pm {g_{\psi\psi,r}\over2g^{(\pm)}_{rr} }\sqrt{-g^{(\pm)}_{\lambda\lambda}g^{(\pm)}_{rr}}
\equiv \dot{\lambda}_\pm B_{\pm}(R, \theta).
\end{align}

\subsection{The second Israel junction condition}
\label{subsec:second_Israel}
We here introduce the Israel junction condition and derive the integral of the equation of motion for the nucleated bubble wall. The second junction conditions are given by
\begin{align}
K_{\theta\theta}^{(+)} - K_{\theta \theta}^{(-)}&=\dot{\lambda}_{+}A_+-\dot{\lambda}_{-}A_-=-8\pi G \left( S_{\theta \theta} - \frac{1}{2} h_{\theta \theta} S \right),\label{junc_th_th} \\
K_{\psi \psi}^{(+)} - K_{\psi \psi}^{(-)}&=\dot{\lambda}_{+}B_+-\dot{\lambda}_{-}B_-=-8\pi G \left( S_{\psi \psi} - \frac{1}{2} h_{\psi \psi} S \right).\label{junc_ps_ps}
\end{align}
The energy momentum tensor of the bubble wall $S_{ab}$ is defined as follows:
\begin{equation}
S_{ab} \equiv \text{diag} (\sigma, \ p_{\theta} h_{\theta \theta}, \ p_{\psi} h_{\psi \psi}),
\end{equation}
where $\sigma$ is the energy density, and $p_{\theta \theta}$ and $p_{\psi \psi}$ are anisotropic pressure. The junction conditions (\ref{junc_th_th}) and (\ref{junc_ps_ps}) reduce to
\begin{align}
&\dot{\lambda}_{+}A_+-\dot{\lambda}_{-}A_-=-4\pi G \left( (p_\theta - p_\psi) + \sigma \right) h_{\theta \theta},\\
&\dot{\lambda}_{+}B_+-\dot{\lambda}_{-}B_-=-4\pi G \left( -(p_\theta - p_\psi) + \sigma \right)h_{\psi \psi},
\end{align}
and eliminating $(p_\theta - p_\psi)$ from the junction conditions, we obtain
\begin{align}
&\dot{\lambda}_+ \Xi_+ - \dot{\lambda}_- \Xi_- = -4 \pi G \sigma \equiv -\Sigma, \label{reduced_junction}\\
&\Xi_{\pm} \equiv \frac{A_{\pm}/h_{\theta \theta} + B_{\pm}/h_{\psi \psi}}{2} = \frac{1}{2} \left( \partial_{r} \log{\sqrt{h_{\theta \theta} h_{\psi \psi}}} \right) \sqrt{- \frac{g_{\lambda \lambda}^{(\pm)}}{g_{rr}^{(\pm)}}}.
\end{align}
Squaring both side of (\ref{reduced_junction}), we obtain the following equation
\begin{eqnarray}
\left[ \frac{(-1-g_{rr}^{(+)}\dot{R}^2) \Xi_+^2}{{g^{(+)}_{\lambda\lambda}}}-\frac{(-1-g_{rr}^{(-)} \dot{R}^2)\Xi_-^2}{g_{\lambda\lambda}^{(-)}} -\Sigma^2\right]^2=\frac{4\Xi_-^2{\Sigma}^2 (-1-g_{rr}^{(-)}\dot{R}^2)}{g_{\lambda\lambda}^{(-)}},
\end{eqnarray}
and using $g_{rr}^{(+)} \Xi_+^2/{g^{(+)}_{\lambda\lambda}}
=g_{rr}^-\Xi_-^2/g_{\lambda\lambda}^{-}$, we finally obtain
\begin{eqnarray}
\dot{R}^2+V(R,\theta)=0,
\end{eqnarray}
where,
\begin{eqnarray}
V(R,\theta) \equiv {1\over g_{rr}^{(-)}}+{g^{(-)}_{\lambda\lambda}\over 4\Sigma^2 \Xi_-^2 g^{(-)}_{rr}} \biggl[ \Xi_+^2/g^{(+)}_{\lambda\lambda}+\Sigma^2-\Xi_-^2/g^{(-)}_{\lambda\lambda}\biggr]^2.
\end{eqnarray}
\begin{figure}[h]
 \centering
    \includegraphics[width=0.85 \textwidth]{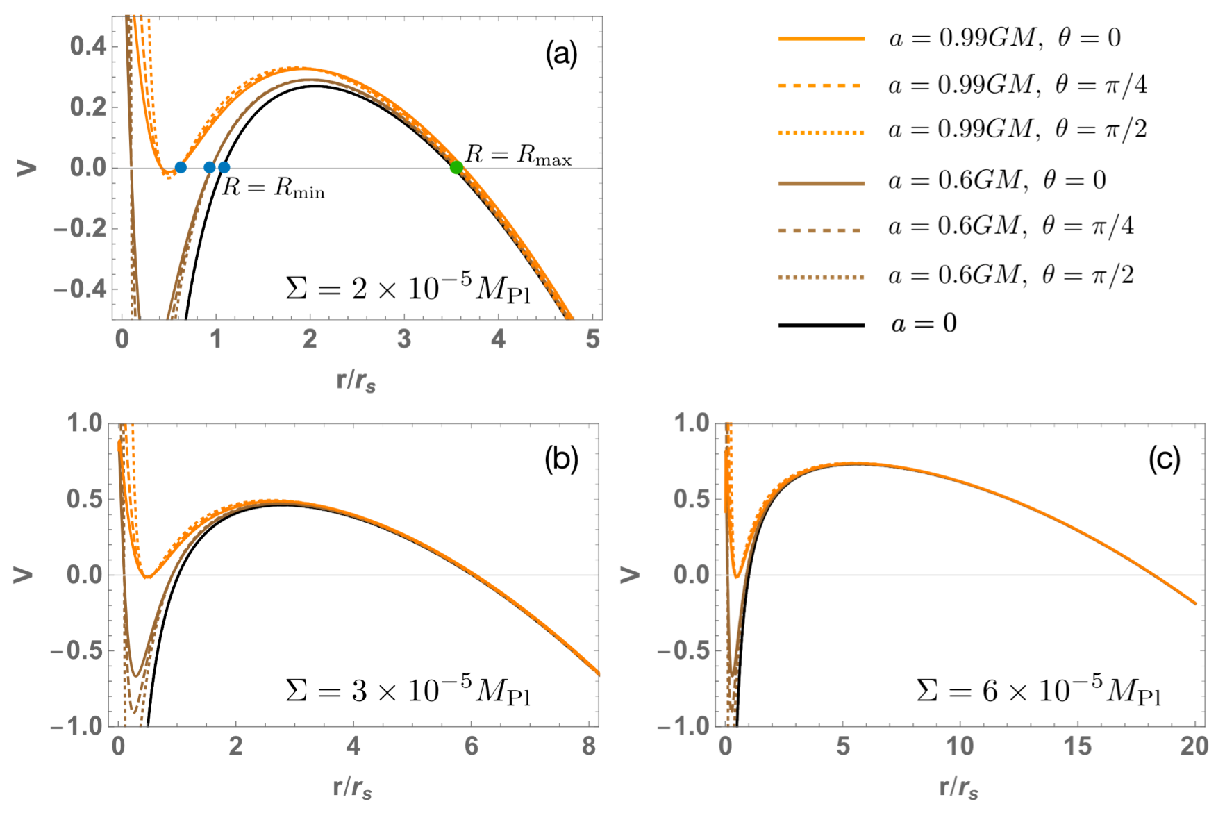}
\caption{The effective potential governing the position of the wall with $a/(GM)=0$, $0.6$, and $0.99$. The blue and green points in (a) show $R=R_{\text{min}}$ and $R = R_{\text{max}}$, respectively. In the Euclidean oscillation phase ($R_{\text{min}}\leq R \leq R_{\text{max}}$), there is less dependence on $\theta$. Here we take $r_s=10^3 \ell_{\text{Pl}}$, $H = 1 \times 10^{-4} M_{\text{Pl}}$, (a) $\Sigma/M_{\text{Pl}} = 2 \times 10^{-5}$, (b)$3 \times 10^{-5}$, and (c) $6 \times 10^{-5}$.
}
\label{potential}
\end{figure}
The effective potential $V(R, \theta)$ governs the position of the wall, and so this determines the dynamics of the wall. In the Euclidean picture the bubble wall oscillates between $R = R_{\text{max}}$ and $R = R_{\text{min}}$, where $R_{\text{max}}$ and $R_{\text{min}}$ are the largest and second largest roots of $V (R, \theta) = 0$, respectively. In order for the assumption that the surface of the shell is characterized by (\ref{surface_shell}) to hold during the whole Euclidean periodic motion, the $\theta$-dependence of the Euclidean motion is not allowed. However, it has its small $\theta$-dependence due to $V = V(R, \theta)$ (FIG. \ref{potential}). We found out that the $\theta$-dependence is relatively larger for a higher spin parameter or for a smaller value of $\Sigma$. We will come back to this point in Sec. \ref{subsec:euclidean_action} and will show the error in the Euclidean action is negligible compared to its spin-dependence.

\subsection{Classification of vacuum bubble solutions and the spin-dependence of the critical mass}
\label{subsec:class}
\begin{figure}[t]
 \centering
    \includegraphics[width=0.9 \textwidth]{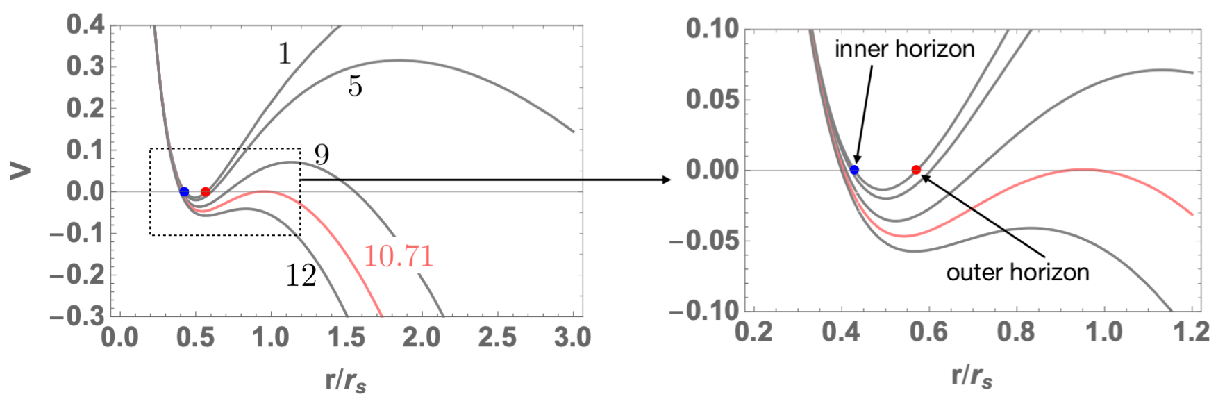}
\caption{Plots of the effective potential with $M_+/M_{\text{Pl}} = 1, 5, 9,$ $10.71$ (critical mass), and $12$. We here take $a=0.99 GM$, $H = 5 \times 10^{-3} M_{\text{Pl}}$ and $\Sigma = 5 \times 10^{-4} M_{\text{Pl}}$. The blue and red points represent the position of the inner and outer horizons, respectively. The pink lines show the effective potential with the critical mass $M_{+} = M_{\text{C}}$.
}
\label{potential_various}
\end{figure}

\begin{figure}[t]
 \centering
    \includegraphics[width=0.7 \textwidth]{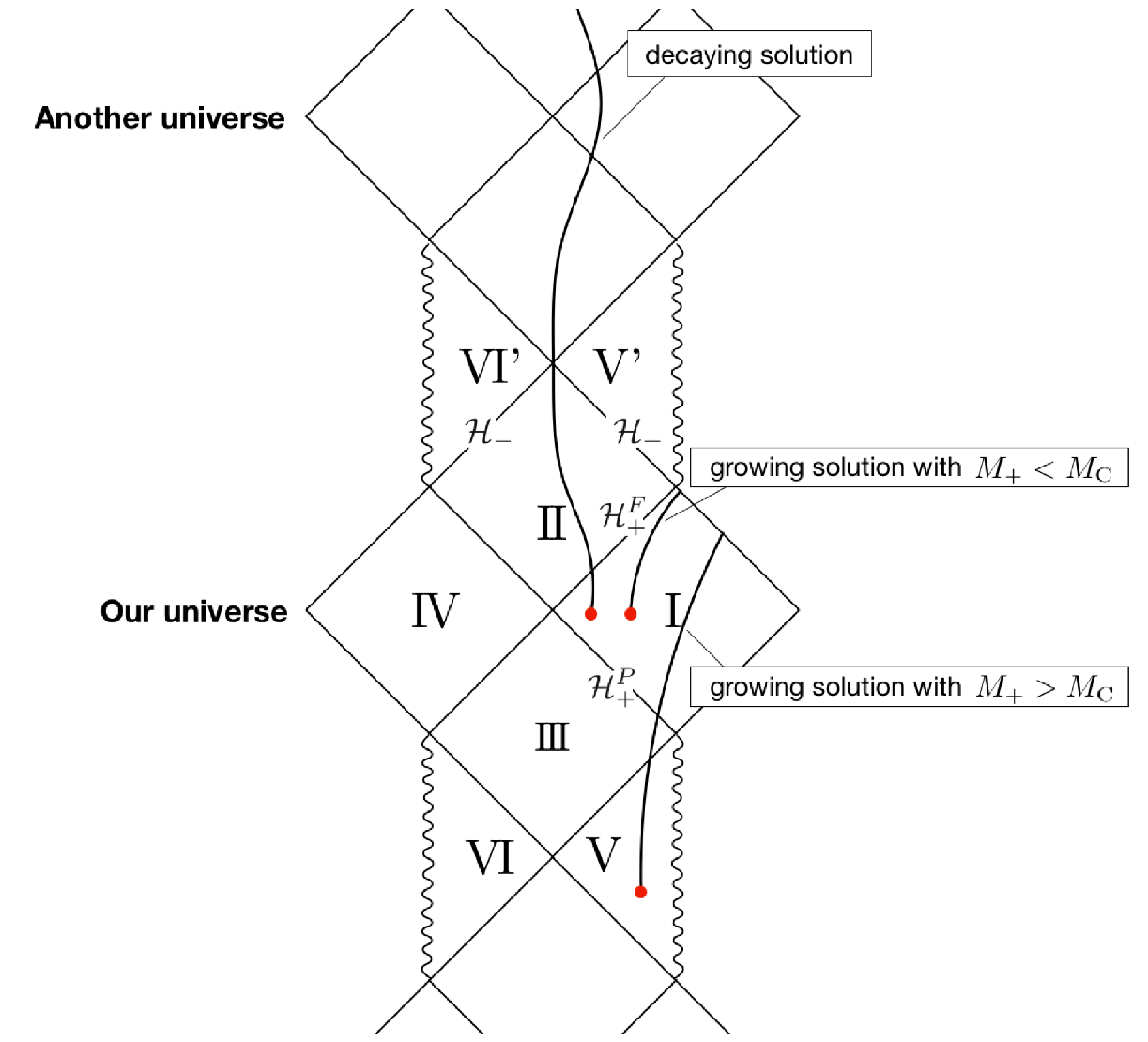}
\caption{The trajectories of the three solutions on the Penrose diagram of Kerr spacetime (black solid lines). The red points show the nucleation points of the vacuum bubbles. Only the growing bubble with $M_+ \leq M_{\text{C}}$ is a physically meaningful solution. ${\cal H}_+^F$ (${\cal H}_+^P$) represents the future (past) outer horizon and ${\cal H}_-$ is the inner horizon.
}
\label{penrose}
\end{figure}

\begin{figure}[t]
\centering
    \includegraphics[width=0.7 \textwidth]{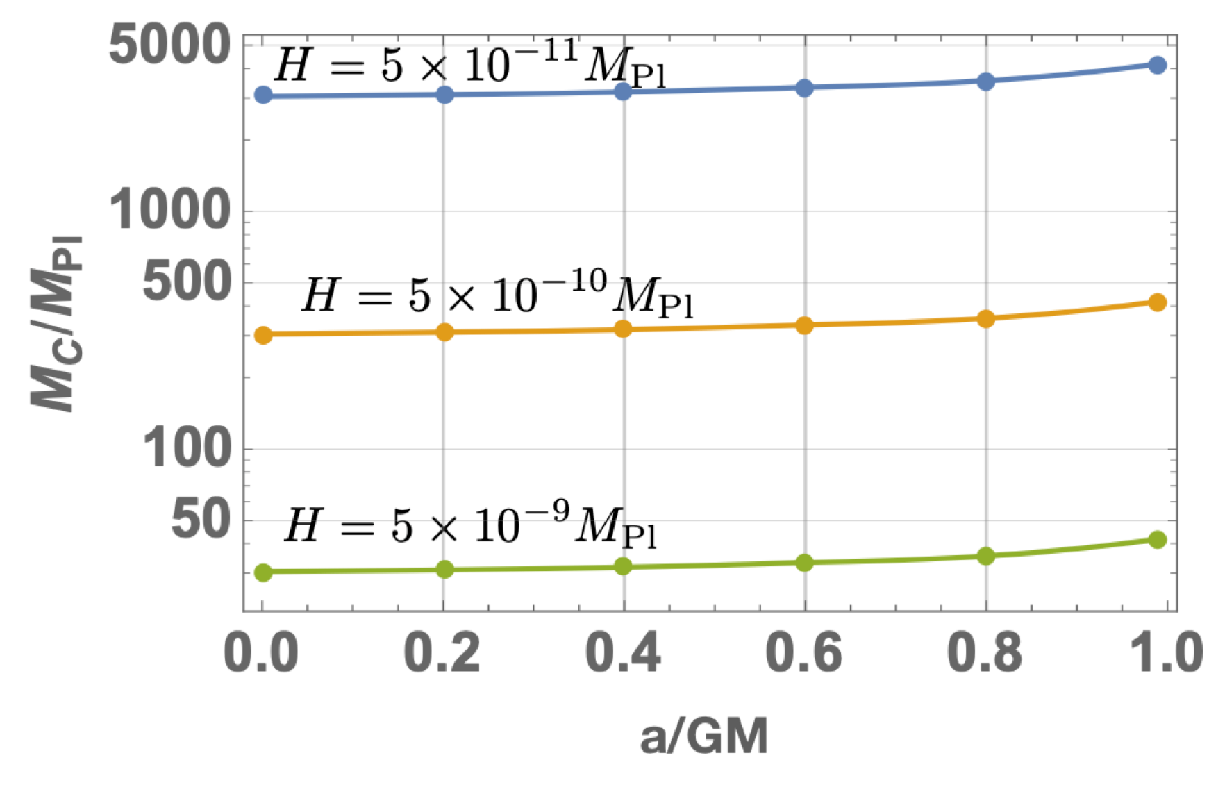}
\caption{Plot of the spin-dependence of the critical mass with different values of $H$ and $\Sigma/H = 4 \times 10^{-7}$.
}
\label{critical}
\end{figure}

It was shown that in order for a non-spinning BH to be a catalyst for vacuum decays, its mass should be smaller than the critical mass, $M_c$, above which there is no $O(3)$-Euclidean solution of a thin-wall vacuum bubble. In this subsection, we investigate the spin-dependence of the critical mass. To this end, let us first classify the solution of thin-wall vacuum bubbles obtained from the Israel junction condition. In Fig. \ref{potential_various}, the effective potentials with various mass parameters ($M_+/M_{\text{Pl}} = 1, 5, 9,$ $10.71$ (critical mass), and $12$) are shown\footnote{We fixed with $\theta = \pi /4$ to plot $V(\theta, R)$, but the qualitative feature in $V$ does not change for any value of $\theta$.}. From Fig. \ref{potential_various}, one can read that the potential has three intersections with the axis of $V=0$ when $M_+ < M_{\text{C}}$ while it has two intersections for $M_+ = M_{\text{C}}$ and one intersection for $M_+ > M_{\text{C}}$.
\begin{figure}[t]
 \centering
    \includegraphics[width=0.9 \textwidth]{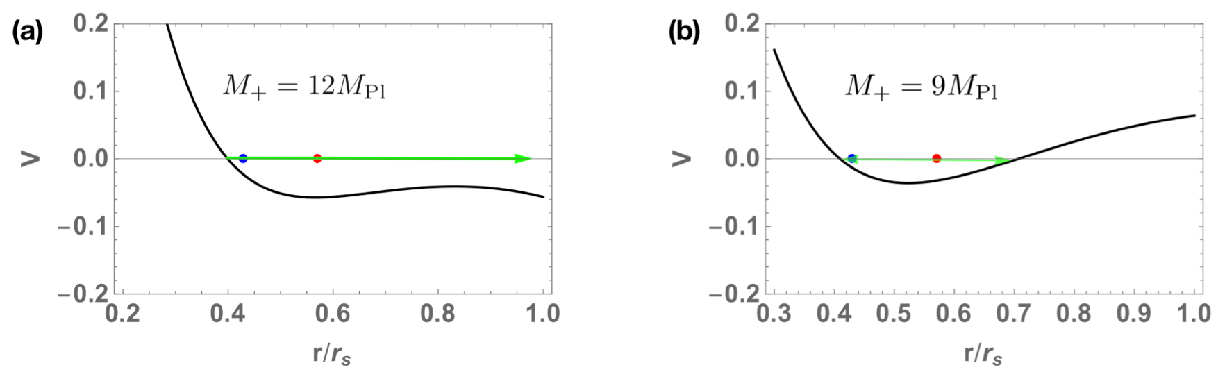}
\caption{(a) shows the trajectory of the growing bubble with $M_+ > M_{\text{C}}$ and (b) shows the oscillatory trajectory of the vacuum bubble trapped around the inner and outer horizons. Both trajectories are depicted with green arrows, and red and blue points represent the position of the outer and inner horizons, respectively.
}
\label{bubble_solutions}
\end{figure}
Therefore, there are two additional solutions other than the growing mode nucleated at the largest radius: one is a growing mode with $M_+ > M_{\text{C}}$, which expands out of the inner horizon (Fig. \ref{bubble_solutions}-(a)) and another solution is an oscillatory solution with $M_+ < M_{\text{C}}$ which crosses the outer and inner horizons repeatedly (Fig. \ref{bubble_solutions}-(b)). However, a vacuum bubble expanding out of the inner horizon is an unphysical picture since the past horizon is assumed to exist in such a solution. Moreover, the oscillatory trajectory between the outside the outer horizon and inside the inner horizon (see Fig. \ref{penrose}) is not realistic since the configuration of the ring singularity would not be stable. This means that the oscillatory solution should be replaced by the decaying solution, in which a vacuum bubble would be nucleated outside the outer horizon and just collapse into the BH singularity without going to another Universe. Therefore, neither of them can be physical/realistic solutions, and only the growing mode with $M_+ \leq M_{\text{C}}$ can be a physically meaningful solution.
Now we found out that the vacuum bubble solution exists only for $M_+ \leq M_{\text{C}}$, and so the critical mass, $M_{\text{C}}$, is a very important quantity to discuss the vacuum decay around a spinning BH. The spin-dependence of $M_{\text{C}}$ is shown in Fig. \ref{critical}. One finds out that $M_{\text{C}}$ increases as the spin parameter $a$ increases and the increment is at most around $30\%$. 

\subsection{Euclidean action and the decay rate around a Kerr BH}
\label{subsec:euclidean_action}
The decay rate of a false vacuum can be estimated by calculating the on-shell Euclidean action when the semi-classical approximation is valid. The decay rate, $\Gamma$, has the form of
\begin{equation}
\Gamma \simeq D_{\text{pre}} e^{-S_{\rm E}},
\end{equation}
where $S_{\rm E}$ is the Euclidean action determined by the classical Euclidean path, and $D_{\text{pre}}$ is a pre-factor that originates from the fluctuations around the Euclidean path. Let us calculate the Euclidean action which consists of both matter and gravity sectors.
To this end, we implement the Wick rotation, $\lambda \to -i \lambda_{\text{E}}$ and $\tau \to -i \tau_{\text{E}}$, and decompose the Euclidean action into four parts as we did in Sec. \ref{sec:review}:
\begin{equation}
S_{\rm E} [\phi, g_{\mu \nu}] = S_{\cal H} + S_+ +S_- +S_{\cal W}.
\end{equation}
The four parts of Euclidean action are given by
\begin{align}
S_{\cal H} &= - \frac{1}{16 \pi G} \int_{\mathcal H} {\mathcal R} + \frac{1}{8\pi G} \int_{\partial \mathcal H} \tilde{K}_{\text{E}},\\
S_{\pm} &= - \frac{1}{16 \pi G} \int_{\mathcal M_{\pm}} {\mathcal R} - \int_{\mathcal M_{\pm}} {\mathcal L}_m + \frac{1}{8\pi G} \int_{\partial \mathcal M_{\pm}} \tilde{K}_{\text{E} \pm},\\
S_{\cal W} &= - \int_{\mathcal W} {\mathcal L}_m = \int_{\mathcal W} \sigma,
\end{align}
where we omitted to write the volume elements in the integrals, and as a reminder,
$\mathcal H$ is the region in the vicinity of the event horizon, ${\mathcal M}_{+}$ (${\mathcal M}_{-}$) is the exterior (interior) region, and $\mathcal W$ is the surface of the bubble. After some calculations (see the Appendix for the computation of $S_{\cal H}$), those terms reduce to \cite{Gregory:2013hja}
\begin{align}
S_{\cal H} &= - \frac{A}{4 G},\\
S_{\pm} & =  \frac{1}{8 \pi G} \int_{\mathcal W} \tilde{K}_{\text{E} \pm} + \frac{1}{8 \pi G} \int_{\mathcal W} \tilde{n}_{\pm \nu} \tilde{u}_{\pm}^{\mu} \nabla_{\mu} \tilde{u}_{\pm}^{\nu},
\end{align}
where $\tilde{n}_{\pm}^{\mu} \equiv \pm n_{\pm}^{\mu}$ and $\tilde{u}^{\mu}$ is the normal vector on the equal-time slice $\Sigma_{\lambda_{\text{E}}}$, which is given by
\begin{equation}
\tilde{u}_{\pm}^{\mu} = \text{diag} (1/\sqrt{g^{(\pm)}_{\lambda_{\text{E}} \lambda_{\text{E}}}},0,0,0).
\end{equation}
\begin{figure}[t]
 \centering
    \includegraphics[width=0.95 \textwidth]{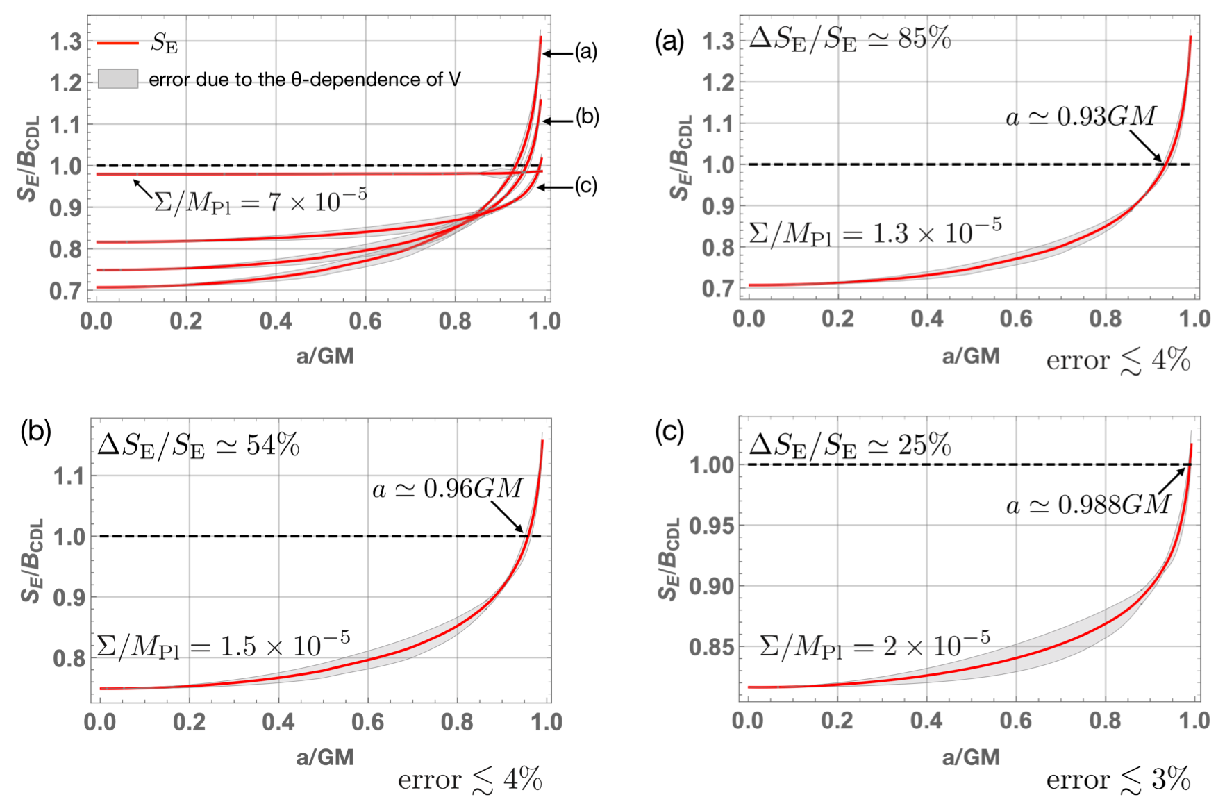}
\caption{The spin-dependence of the Euclidean action $S_{\rm E} = S_{\rm E} (a)$ while the parameters are fixed as $r_{s} = 10^3 \ell_{\text{Pl}}$, $H = 1 \times 10^{-4} M_{\text{Pl}}$, (a) $\Sigma/M_{\text{Pl}} = 1.3 \times 10^{-5}$, (b) $1.5 \times 10^{-5}$, and (c) $2 \times 10^{-5}$. The black dashed lines show $S_{\text{E}} = B_{\text{CDL}}$. Here we defined $\Delta S_{\text{E}} / S_{\text{E}} \equiv |S_{\text{E}} (a=0) - S_{\text{E}} (a=0.99GM)|/ S_{\text{E}} (a=0)$. For the comparison, the action for $\Sigma = 7 \times 10^{-5} M_{\text{Pl}}$ is also shown in the upper left figure.
}
\label{action}
\end{figure}
The surface gravity on the wall $\kappa_{\pm} \equiv \tilde{n}_{\pm \nu} \tilde{u}_{\pm}^{\mu} \nabla_{\mu} \tilde{u}_{\pm}^{\nu}$ is therefore given by
\begin{equation}
\kappa_{\pm} = \pm \Gamma^{r}_{\pm \lambda_{\text{E}} \lambda_{\text{E}}} \sqrt{g_{rr}^{(\pm)}/g_{\lambda_{\text{E}} \lambda_{\text{E}}}^{(\pm)}} \dot{\lambda}_{\text{E} \pm} \equiv \pm \Theta_{\pm} \dot{\lambda}_{\text{E} \pm},
\end{equation}
where a dot denotes the derivative with respect to the Euclidean proper time $\tau_E$. Now we have the explicit form of the exponent factor $S_{\rm E}$ as
\begin{align}
\begin{split}
S_{\rm E} &= - \frac{\Delta A}{4G} - \frac{1}{2} \int_{\mathcal W} \sigma + \frac{1}{8 \pi G} \int_{\mathcal W} (\kappa_+ + \kappa_-),\\
&= - \frac{\Delta A}{4G} - \frac{1}{8 \pi G} \int_{\mathcal W} \left\{ \left( \Xi_- + \Theta_- \right) \dot{\lambda}_{\text{E} -}
- \left( \Xi_+ + \Theta_+ \right) \dot{\lambda}_{\text{E} +} \right\},\\
& = - \frac{\Delta A}{4G} - \frac{1}{8 \pi G} \oint d\tau_{E} d\theta d\psi \sqrt{h_{\text{E}}} \left[ \sqrt{g_{\lambda_{\text{E}} \lambda_{\text{E}}}^{(-)}/g_{rr}^{(-)}} \left( \frac{1}{2} \partial_r \ln{\sqrt{g_{\theta \theta} g_{\psi \psi}}} + \frac{g_{rr}^{(-)}}{g_{\lambda_{\text{E}} \lambda_{\text{E}}}^{(-)}} \Gamma^r_{- \lambda_{\text{E}} \lambda_{\text{E}}} \right) \dot{\lambda}_{\text{E} -}
\right.\\
&\left.
~~~~~~~~~~~~~~~~~~~~~~~~~~~~~~~~~~~~~~~
- \sqrt{g_{\lambda_{\text{E}} \lambda_{\text{E}}}^{(+)}/g_{rr}^{(+)}} \left( \frac{1}{2} \partial_r \ln{\sqrt{g_{\theta \theta} g_{\psi \psi}}} + \frac{g_{rr}^{(+)}}{g_{\lambda_{\text{E}} \lambda_{\text{E}}}^{(+)}} \Gamma^r_{+ \lambda_{\text{E}} \lambda_{\text{E}}} \right) \dot{\lambda}_{\text{E} +} \right],
\end{split}
\end{align}
where $\Gamma^{r}_{\pm \lambda_{\text{E}} \lambda_{\text{E}}} = - \partial_r g_{\lambda_{\text{E}} \lambda_{\text{E}}}^{(\pm)}/(2 g_{rr}^{(\pm)})$, $\Delta A$ is the change of horizon area before and after the vacuum decay, and we assume $(p_{\theta} + p_{\psi})/2 = - \sigma$. We numerically calculated the Euclidean action and compared it to the CDL Euclidean action $B_{\text{CDL}}$ that is given by \cite{Coleman:1980aw}
\begin{equation}
B_{\text{CDL}} = \frac{\pi}{GH^2} \frac{\Sigma^4 /H^4}{(1-\Sigma^2/H^2)^2}.
\end{equation}
We found that the action for $a>0$ is larger than that for the non-spinning BH. For instance, when $a =0.99 GM$ and $\Sigma = 1.3 \times 10^{-5} M_{\text{Pl}}$, it is larger than $\sim 85\%$ of the action for $a=0$ (Fig. \ref{action}), which is greater than the error due to the $\theta$-dependence of the potential $V(R, \theta)$ (see the gray shaded region in Fig. \ref{action}). Remarkably, depending on the value of $\Sigma$, the action for a highly spinning BH ($a \gtrsim 0.9$) is greater than the CDL action. Therefore, we can conclude that the Euclidean action of the ellipsoidal vacuum bubble is larger than that of the spherical bubble around a non-spinning BH, and it is also found out that in the existence of a highly spinning BH, a false vacuum state can be more stabilized than expected from the CDL solution, provided that the vacuum bubble is the ellipsoid of (\ref{surface_shell}). However, a dense bubble wall results in the less spin-dependence of the action and the catalyst effect is also suppressed (see the upper left figure of Fig. \ref{action}).

\section{Conclusion}
We calculated the Euclidean action of an ellipsoidal vacuum bubble around a spinning BH, whose major and minor axes are given by $\sqrt{a^2 + R^2}$ and $R$, respectively. As an example, we consider a vacuum decay from the zero-energy vacuum state to an AdS vacuum, but our analysis can be easily extended to other cases of vacuum decay. The primary assumption is that a nucleated vacuum bubble has its thin wall, and so we used the Israel junction condition to investigate the Euclidean dynamics of the bubble. We assumed that the typical vacuum bubble has its time-like surface determined by the angular metric $g_{\psi \psi}^{(\pm)}|_{r_{\pm} = R}$ and $g_{\theta \theta}^{(\pm)}|_{r_{\pm} = R}$, which is a natural extension of the case of vacuum decay around a non-rotating BH \cite{Gregory:2013hja}. In order for the first Israel junction condition to be satisfied, our analysis was restricted to the case of vacuum decay by which only small-change of vacuum energy is involved, $a^2 \ll l^2$, and the spin and mass of the BH does not change, i.e., $a_+ = a_-$ and $M_+ = M_-$. We derived the second Israel junction condition that reduces to the integral of the equation of motion of the thin wall, based on which we investigated the Euclidean motion of the wall. We found three kinds of bubble solutions, but only the growing bubble solution with $M_+ < M_{\text{C}}$ is physical and realistic. The other two solutions cannot be realistic unless the past horizon exists or the ring singularity is stable. The critical mass $M_{\text{C}}$ increases due to the spin effect, which means that the mass range of a spinning BH, in which the vacuum decay can occur, is larger than that of a non-spinning BH.
We found out that there exists small $\theta$-dependence in the Euclidean motion, which might be inconsistent with our earlier assumption that the surface of the wall is characterized by $r_{\pm} = R (\lambda_{\pm})$. The $\theta$-dependence is relatively larger for a highly spinning BH but it is suppressed for a larger $\Sigma$. The increment of the Euclidean action due to the spin of BH overwhelms the error (see Fig. \ref{action}). Therefore, we can conclude that the spin effect of BH can suppress the vacuum decay rate compared to that for a non-spinning case, provided that the ellipsoidal vacuum bubble characterized by (\ref{surface_shell}) gives the least action. It is also found out that in the existence of a near-extremal black hole, a false vacuum state can be more stabilized than the case of the CDL solution.
Note that whatever the change in the determinant $D_{\text{pre}}$ is, it is not likely to be comparable to the change in $e^{-S_{\text{E}}}$.

The vacuum decay around BHs are very important phenomenon from the point of view of particle physics and early cosmology. Recently, the vacuum decay around a non-spinning BH was applied to put constraints on the abundance of non-spinning primordial BHs (PBHs), whose mass is comparable to or smaller than $10^6 M_{\text{Pl}}$. It can be translated into the constraints on the parameters of Higgs potential and on the spectral index of density fluctuations \cite{Dai:2019eei}. Our result shows that a BH with a smaller spin more strongly enhances the decay rate compared to the case of a highly spinning BH. If the vacuum decay rate around a spinning BH was much higher than that around a non-spinning BH, it would be necessary to seriously take into account sub-dominant spinning PBHs (see e.g. \cite{He:2019cdb,Harada:2017fjm}) to put constraint on the PBH abundance. In this sense, our result provides a supporting evidence to guarantee that taking only non-spinning PBHs into account may be enough to put constraint on the PBH abundance and on the parameters of the Higgs potential.

Throughout this paper, we restricted ourselves to the vacuum decay around a four-dimensional non-charged spinning BH. An interesting generalization of our study would be to explore vacuum decays in the Kerr-Newman background or higher-dimensional Kerr background, which may lead to new interesting black hole phenomenology.

\appendix
\section{Conical deficit regularization for an axisymmetric Euclidean metric}

In this appendix we present the calculation of $S_{\cal H}$ with a $\theta$-dependent Euclidean metric since the Kerr metric has its axisymmetric structure. This is a generalization of the computation presented in the Appendix A in \cite{Gregory:2013hja} where the spherical symmetry of the Euclidean metric is assumed. We assume that the conical singularity with its deficit angle $\delta$ can be locally parameterized by cylindrical coordinates, $\{ (z, \chi) | \ 0\leq z, \ 0 \leq \chi \leq 2 \pi \}$, and the transverse space is independent of the cylindrical structure near the singularity $z \to 0$, that is, the metric has the form
\begin{equation}
ds^2 = dz^2 + B^2(z, \theta) d\chi^2 + C^2 (z,\theta) d\theta^2 + D^2(z, \theta) d\phi^2,
\end{equation}
with $\partial_z C(0, \theta) = \partial_z D(0, \theta) = 0$. The main idea in \cite{Gregory:2013hja} is to smooth out the conical singularity by taking a function $B$ such that $\partial_z B (0,\theta) = 1$ and $\partial_z B (\epsilon,\theta) = 1-\delta (\theta)$. Then we can take the limit of $\epsilon \to 0$ at the final stage of the calculation. Computing the Ricci scalar ${\cal R}$ near the conical singularity, one obtains
\begin{align}
\begin{split}
&{\cal R} = -\frac{2}{B C^3 D} \left[ - B \partial_{\theta} C \partial_{\theta} D
+ C (\partial_{\theta} B \partial_{\theta} D+ B \partial_{\theta}^2 D)
+ B C^2 \partial_z C \partial_z D \right.\\
&\left. +D (-\partial_{\theta} B \partial_{\theta} C + C \partial_{\theta}^2 B + C^2 (\partial_z B \partial_z C + C \partial_z^2 B + B \partial_z^2 C)) + C^3 (\partial_z B \partial_z D + B \partial_z^2 D) \right].
\end{split}
\label{app:ricci}
\end{align}
In (\ref{app:ricci}) there is an unbounded term of $\partial_z^2 B \simeq (\partial_z B(\epsilon,\theta) - \partial_z B(0,\theta))/\epsilon = -\delta / \epsilon$ with respect to the limit of $\epsilon \to 0$. Other terms are regular and so the following integration of the Ricci scalar reduces to
\begin{align}
\begin{split}\displaystyle
&-\frac{1}{16 \pi G} \lim_{\epsilon \to +0} \int_{\cal H} d^4 x \sqrt{B^2 C^2 D^2} {\cal R}\\
&=-\frac{2\pi}{16 \pi G} \lim_{\epsilon \to +0} \int_0^{\epsilon} dz \int d\theta d \phi B C D \frac{-2 \partial_z^2 B}{B} =- \frac{1}{4 G} \int d\theta d\phi CD \delta (\theta).
\end{split}
\label{app:ricci2}
\end{align}
Next we compute the GHY term on the boundary $\partial{\cal H}$ with the extrinsic curvature $K = \nabla_{\mu} n^{\mu}_H = - \partial_z \log{(BCD)}$, where the inward pointing normal vector is given by $n_H^{\mu} = - \delta_{z \mu}$.
Hence the GHY term is
\begin{align}
\begin{split}
&\frac{1}{8 \pi G} \int_{\partial {\cal H}} d\chi d\theta d \phi \sqrt{B^2C^2D^2} K\\
&=-\frac{1}{4 G} \int_{z = \epsilon} d\theta d \phi \sqrt{B^2C^2D^2} \frac{\partial_z B}{B} = - \frac{1}{4G} \int d\theta d\phi CD (1-\delta (\theta)).
\label{app:GHY}
\end{split}
\end{align}
From (\ref{app:ricci2}) and (\ref{app:GHY}), one finally obtains the conical deficit action
\begin{align}\displaystyle
\begin{split}
S_{\cal H} &= -\frac{1}{16 \pi G} \lim_{\epsilon \to +0} \int_{\cal H} d^4 x \sqrt{B^2 C^2 D^2} {\cal R} + \frac{1}{8 \pi G} \int_{\partial {\cal H}} d\chi d\theta d \phi \sqrt{B^2C^2D^2} K\\
&= - \frac{1}{4 G} \int d\theta d\phi CD = - \frac{A}{4G},
\end{split}
\end{align}
where ${A} \equiv \int d\theta d\phi CD$ is the horizon area.

\acknowledgments

This work was supported by the Perimeter Institute for Theoretical Physics, JSPS Overseas Research Fellowships (N.O.), Sasakawa Scientific Research Grant from the Japan Science Society (K.U.), JSPS Grant-in-Aid for Scientific Research Nos.~18K18764 (M.Y.), MEXT KAKENHI Grant-in-Aid for Scientific Research on Innovative Areas Nos.~15H05888, 18H04579 (M.Y.), Mitsubishi Foundation (M.Y.), JSPS and NRF under the Japan-Korea Basic Scientific Cooperation Program (M.Y.), and JSPS Bilateral Open Partnership Joint Research Projects (M.Y.).
Research at the Perimeter Institute is supported by the Government of Canada through Industry Canada, and by the Province of Ontario through the Ministry of Research and Innovation.

\end{document}